\documentclass{ws-ijmpcs}

\begin{document}

\markboth{S.~P.~Kim}
{Quantum Dynamics for de Sitter Radiation}

\catchline{}{}{}{}{}
%
%%%%%%%%%%%%%%%%%%%%%%%%%%%%%%%%%%%%%%%%%%%%%%%%%%%%%%%%%%%%%%%%%%%%

\title{QUANTUM DYNAMICS FOR DE SITTER RADIATION}

\author{SANG PYO KIM}

\address{Department of Physics, Kunsan National University,
Kunsan 573-701, Korea\\Institute of Astrophysics, Center for Theoretical Physics, Department of Physics, National Taiwan University,Taipei 106, Taiwan\\
sangkim@kunsan.ac.kr}

\maketitle

\begin{history}
\received{Day Month Year}
\revised{Day Month Year}
\end{history}

\begin{abstract}
We revisit the Hamiltonian formalism for a massive scalar field and study the particle production in a de Sitter space.
In the invariant-operator picture the time-dependent annihilation and creation operators are constructed
in terms of a complex solution to the classical equation of motion for the field and
the Gaussian wave function for each Fourier mode is found which is an exact solution to the Schr\"{o}dinger equation.
The in-out formalism is reformulated by the annihilation and creation operators
and the Gaussian wave functions. The de Sitter radiation from the in-out formalism
differs from the Gibbons-Hawking radiation in the planar coordinates, and we discuss the discrepancy of the
particle production by the two methods.

\end{abstract}

\ccode{PACS numbers: 04.06.-m; 04.62.+v; 11.10.Ef; 98.80.Cq}

\section{Introduction}

The universe with a cosmological constant is the pure de Sitter (dS) space and
has the maximal symmetry that makes a quantum field theory more tractable than other curved spacetimes
(for a review and references, see ref.~[\refcite{birrell-davies}]).
In spite of numerous works since the discovery of the dS-invariant vacuum
by Bunch and Davies\cite{bunch-davies} and the dS radiation by Gibbons and Hawking\cite{gibbons-hawking}
(for instance, see ref.~[\refcite{mottola}]), the proper selection of the physical vacuum of
a quantum field in dS space has recently been challenged by Polyakov.\cite{polyakov}
It has been further argued that the cosmological constant problem may be resolved by IR physics
through the UV/IR mixing from an interaction.\cite{krotov-polyakov}

In this paper, we revisit the quantum field theory for a massive scalar field in the dS space
by unifying the in-out formalism by Schwinger and DeWitt\cite{schwinger,dewitt}
and the invariant-operator picture by Lewis and Riesenfeld.\cite{lewis-riesenfeld}
The in-out formalism provides a good framework for quantum field theory involving particle production.
It has been known that an expanding spacetime produces particles\cite{parker} and that
the dS space has the Gibbons-Hawking radiation\cite{gibbons-hawking,mottola} (for a recent
discussion, see ref. [\refcite{haro-elizalde}]). Recently the dS radiation has
 been interpreted in connection with the Schwinger
mechanism.\cite{polyakov,krotov-polyakov,kim2007,akhmedov}
In contrast to the early works on the in-out formalism,\cite{kim2010}
in this paper we shall directly use the wave functions for the field and shall find
the vacuum persistence amplitude between two spacelike hypersurfaces, one in the past infinity
and the other at an arbitrary time. In cosmological scenarios
the Gaussian wave functions on each hypersurface carry the same information as the Heisenberg field operator.
In fact, the in-out formalism and the invariant-operator picture formalism are equivalent to each other.

In the Hamiltonian formalism a massive scalar field in the dS space is equivalent to an infinite sum of time-dependent
harmonic oscillators.\cite{kim1995} The invariant-operator picture provides a pair of time-dependent annihilation and
creation operators that generate all the exact quantum states in analogy with a
harmonic oscillator.\cite{mmt,bfv,kim-page,kim-kim,pfr,gds} Using these
operators we construct the Gaussian wave packets for each oscillator that evolve from the spacelike hypersurface
in the past infinity toward to the future infinity.\cite{bfv,kim-page,kim-kim,pfr,gds} Furthermore,
we find the Bogoliubov transformation between the operators on the two different hypersurfaces.

Interestingly, in the planar coordinates of dS space the number of in-vacuum particles
carried by the Gaussian wave packet on each hypersurface exponentially increases in proportional
to the cube of the scale
factor and the vacuum persistence amplitude confirms this phenomenon.
This contrasts with the Gibbons-Hawking radiation which is a thermal and scale-factor free
distribution in the future infinity. It is shown that the matrix amplitude between the Gaussian wave packet
and the out-vacuum Gaussian wave packet in the future infinity, however,
exhibits the Gibbons-Hawking radiation. The Gibbons-Hawking radiation is
the particle number measured by any detector whose quantum states are prescribed by
the out-vacuum solution. Hence, from the view of the in-out formalism this discrepancy raises
a fundamental question on the quantum field in the planar coordinates of the dS space.

%The organization of this paper is as follows. In section 2 we revisit
%the Hamiltonian formalism for a massive scalar field in an FRW universe. In section 3
%the dS radiation is rederived and in section 4 Gaussian wave packets are reconstructed
%in the invariant operator picture.
%In section 5 the number operator is constructed in terms of the time-dependent
%creation and annihilation operators and in section 6 the number of the in-vacuum particles
%carried by the Gaussian wave function is computed.
%In section 7 we discuss the vacuum persistence amplitude and the vacuum instability and in section 8
%we discuss the Gaussian wave packet for the out-vacuum.

\section{Hamiltonian Formalism for Massive Scalar Field} \label{hamiltonian form}

In a Friedmann-Robertson-Walker spacetime [in the units of $c = \hbar = 1$]
\begin{eqnarray}
ds^2 = dt^2 - a^2 (t) d{\bf x}^2,
\end{eqnarray}
a massive scalar field $\Phi$ has the action
\begin{eqnarray}
S = \int dt L =  \int dt \int d^3 {\bf x} \sqrt{-g} \frac{1}{2} \Bigl(g^{\mu \nu} \partial_{\mu} \Phi \partial_{\nu} \Phi
- m^2 \Phi^2 \Bigr).
\end{eqnarray}
In the Hamiltonian formalism the action with $\Pi = a^3 \dot{\Phi}$
\begin{eqnarray}
S = \int dt (\dot{\Phi} \Pi -  H) ,
\end{eqnarray}
is given by the Hamiltonian and its density
\begin{eqnarray}
H(t) = \int d^3 {\bf x}  \frac{1}{2} \Bigl( \frac{\Pi^2}{a^3} + \frac{(\nabla \Phi)^2}{a^2} + m^2 \Phi^2 \Bigr)
= \int d^3 {\bf x} {\cal H} (t,x). \label{ham0}
\end{eqnarray}
The stress-energy-momentum tensor
\begin{eqnarray}
T_{\mu \nu} = \frac{2}{\sqrt{-g}} \frac{\delta S}{\delta g^{\mu \nu}} = \partial_{\mu} \Phi \partial_{\nu} \Phi
- \frac{1}{2} g_{\mu \rho} \Bigl(g^{\rho \nu} \partial_{\rho} \Phi \partial_{\sigma} \Phi
- m^2 \Phi^2 \Bigr),
\end{eqnarray}
has also the Hamiltonian expression
\begin{eqnarray}
T_{00} &=&  \frac{{\cal H}}{a^3}, \quad T_{0i} = \frac{\Pi}{a^3}  \partial_i \Phi, \nonumber\\
T_{ij} &=&  \frac{a^2}{2} \delta_{ij} \Bigl(\frac{\Pi^2}{a^6} -
 m^2 \Phi^2 \Bigr) + \Bigl( \partial_i \Phi \partial_j \Phi-  \frac{1}{2} \delta_{ij}(\nabla \Phi)^2 \Bigr). \label{stress}
\end{eqnarray}

Now we decompose the Hamiltonian (\ref{ham0}) by the Fourier expansion of the field and the conjugate momentum
\begin{eqnarray}
\Phi (t, {\bf x}) = \int \frac{d^3 {\bf k}} {(2 \pi)^{3/2}} \phi_{\bf k} (t) e^{i {\bf k} \cdot {\bf x}}, \quad \Pi (t, {\bf x}) = \int \frac{d^3 {\bf k}} {(2 \pi)^{3/2}} \pi_{\bf k} (t) e^{i {\bf k} \cdot {\bf x}}. \label{fourier1}
\end{eqnarray}
Then the Hamiltonian is an infinite sum of time-dependent oscillators
\begin{eqnarray}
H (t) = \int d^3 {\bf k} \frac{1}{2} \Bigl( \frac{\dot{\pi}_k^2}{a^3} +  a^3 \omega_k^2 \phi_k^2 \Bigr)
= \int d^3 {\bf k} H_k (t), \label{ham}
\end{eqnarray}
where
\begin{eqnarray}
\pi_k = a^3 \dot{\phi}_k, \quad \omega_k^2 (t) = \frac{k^2}{a^2} + m^2.
\end{eqnarray}
Here and hereafter $\phi_k^2$ and $\pi_k^2$ denote
\begin{eqnarray}
\phi_{k}^2  :&=& \phi_{\bf k} \phi_{- {\bf k}} = \Bigl( \frac{\phi_{\bf k} + \phi_{- {\bf k}} }{2} \Bigr)^2 + \Bigl( \frac{\phi_{\bf k} - \phi_{- {\bf k}} }{2i} \Bigr)^2, \nonumber\\
\pi_{k}^2  :&=& \pi_{\bf k} \pi_{- {\bf k}} = \Bigl( \frac{\pi_{\bf k} + \pi_{- {\bf k}} }{2} \Bigr)^2 + \Bigl( \frac{\pi_{\bf k} - \pi_{- {\bf k}} }{2i} \Bigr)^2, \label{fourier2}
\end{eqnarray}
and $k^2 = {\bf k}^2$. Thus, the $k$-mode denotes both the cosine and the sine modes.
The Hamiltonian (\ref{ham}) has the meaning of $H_k(t) = H_k (\pi_k, \phi_k, \Sigma_t)$ on each spacelike hypersurface $\Sigma_t$
and the Hamiltonian formalism describes the evolution from one hypersurface $\Sigma_{t_0}$ to another $\Sigma_t$.
Similarly, the stress-energy-momentum tensor (\ref{stress}) can be decomposed into the Fourier modes according to eq. (\ref{fourier2}).
The Hamilton equation for $H_k (t)$ is the Fourier mode of the field equation
\begin{eqnarray}
\ddot{\phi}_k + 3 \frac{\dot{a}}{a} \dot{\phi}_k + \omega_k^2 (t) \phi_k = 0. \label{mod eq}
\end{eqnarray}

\section{de Sitter Radiation of Massive Particles in Planar Coordinates} \label{dS radiation}

The maximal symmetry of a dS space allows one to explicitly solve the field equation
for a massive or massless scalar field. In fact, in the planar coordinates, $a(t) = e^{Ht}$,
\begin{eqnarray}
ds^2 = - dt^2 + e^{2Ht} d{\bf x}^2, \label{plan coor}
\end{eqnarray}
eq. (\ref{mod eq}) in the canonical form takes the form of the Morse potential while in the global coordinates, $a (t) = \cosh(Ht)/H$,
its is related to the P\"{o}sch-Teller potential.
The general solution to eq. (\ref{mod eq}) in the metric (\ref{plan coor}) is given by
\begin{eqnarray}
\varphi_k (t) = \Bigl(\frac{\pi}{4H} \Bigr)^{1/2} e^{- 3 Ht/2} \Bigl[  c_{k}^{(1)} e^{-\pi p/2} H_{i p}^{(1)} (z)
+ c_{k}^{(2)} e^{\pi p} H_{i p}^{(2)} (z) \Bigr],
\label{gen sol}
\end{eqnarray}
where $H_{i p}^{(1)}$ and $ H_{i p}^{(2)}$ are the Hankel functions\cite{gr-table} and for massive particles
($m/H > 3/2$)
\begin{eqnarray}
z =  \frac{k}{H} e^{- Ht}, \quad p = \Bigl(\frac{m^2}{H^2} - \frac{9}{4} \Bigr)^{1/2}. \label{p}
\end{eqnarray}
The standard quantization rule requires that the Wronskian condition should be satisfied\cite{dewitt}
\begin{eqnarray}
a^3 (t) {\rm Wr} [\varphi_k (t), \varphi_k^* (t)] = a^3 (t)
\Bigl(\varphi_k (t) \dot{\varphi}_k^* (t) - \dot{\varphi}_k (t) \varphi_k^* (t) \Bigr) =  i.
\label{quant rule}
\end{eqnarray}
Though the integration constants $c_{k}^{(1)}$ and $c_{k}^{(2)}$ can be arbitrary complex numbers,
the quantization rule (\ref{quant rule}) restricts them to satisfy
\begin{eqnarray}
|c_{k}^{(1)}|^2 - |c_{k}^{(2)}|^2 =1.
\end{eqnarray}

In the past infinity $(t = - \infty)$, the general solution (\ref{gen sol}) has the asymptotic form
\begin{eqnarray}
\varphi_k (t) =  c_{k}^{(1)} \varphi_k^{\rm (in)} (t)
+ c_{k}^{(2)} \varphi_k^{{\rm (in)}*} (t),
\label{asym sol1}
\end{eqnarray}
where the in-vacuum (the Bunch-Davies vacuum\cite{bunch-davies} with $c_{k}^{(1)} =1$ and $c_{k}^{(2)} = 0$) is provided by the positive frequency solution with respect to the energy operator, $i \partial_t$,
\begin{eqnarray}
\varphi_k^{\rm (in)} (t) = \frac{e^{-Ht}}{\sqrt{2k}} \exp \Bigl(i \frac{k}{H} e^{- Ht} \Bigr). \label{in sol}
\end{eqnarray}
In the future infinity $(t = \infty)$, using the asymptotic formulae for the Hankel functions\cite{gr-table}
\begin{eqnarray}
H_{i p}^{(1)} (z) &=& J_{i p} (z) + i N_{i p} (z) \nonumber\\
&=& \frac{1}{\sinh (\pi p)} \Bigl[\frac{e^{\pi p}}{\Gamma (1 + i p)} \Bigl(
 \frac{z}{2} \Bigr)^{i p} - \frac{1}{\Gamma (1 - i p)} \Bigl(
 \frac{z}{2} \Bigr)^{- i p}\Bigr], \nonumber\\
H_{i p}^{(2)} (z) &=& J_{i p} (z) - i N_{i p} (z) \nonumber\\ &=& - \frac{1}{\sinh (\pi p)}
\Bigl[\frac{e^{- \pi p}}{\Gamma (1 + i p)} \Bigl(
 \frac{z}{2} \Bigr)^{i p} + \frac{1}{\Gamma (1 - i p)} \Bigl(
 \frac{z}{2} \Bigr)^{- i p}\Bigr],
\end{eqnarray}
the solution (\ref{gen sol}) asymptotically becomes
\begin{eqnarray}
\varphi_k (t) = \alpha_{k} \varphi_k^{\rm (out)} (t)
+ \beta_{k} \varphi_k^{{\rm (out)}*} (t).
\label{asym sol2}
\end{eqnarray}
where the out-vacuum is constructed by another positive frequency solution
\begin{eqnarray}
\varphi_k^{\rm (out)} (t) = \frac{e^{-3Ht/2}}{\sqrt{2 H p}} e^{- i H p t}.
\end{eqnarray}
Here the Bogoliubov coefficients are given by
\begin{eqnarray}
\alpha_{k} &=& \Bigl(\frac{\pi p}{2} \Bigr)^{1/2} \Bigl( c_{k}^{(1)} e^{\pi p/2} + i c_{k}^{(1)} e^{- \pi p/2} \Bigr) \frac{(k/2H)^{i p}}{ \sinh (\pi p) \Gamma (1 + i p)}, \nonumber\\
\beta_{k} &=& - \Bigl(\frac{\pi p}{2} \Bigr)^{1/2} \Bigl( c_{k}^{(1)} e^{- \pi p/2} + i c_{k}^{(1)} e^{\pi p/2} \Bigr) \frac{ (k/2H)^{- i p}}{ \sinh (\pi p) \Gamma (1 - i p)}. \label{bog coef1}
\end{eqnarray}
A direct calculation shows that the Bogoliubov relation holds
\begin{eqnarray}
|\alpha_{k}|^2 - |\beta_{k}|^2 = \Bigl(|c_{k}^{(1)}|^2 - |c_{k}^{(2)}|^2 \Bigr) = 1.
\end{eqnarray}

Noting that $|\beta_k|^2$ is the particle number, the probability for particle production is
\begin{eqnarray}
P_k = \frac{1}{e^{2 \pi p} -1} |c_{k}^{(1)}|^2 + \frac{e^{2 \pi p}}{e^{2 \pi p} -1} |c_{k}^{(2)}|^2.
\label{mean num}
\end{eqnarray}
In the above we have assumed the real $c_{k}^{(1)}$ and $c_{k}^{(2)}$ for simplicity which corresponds to
the zero squeezing angle. The interpretation of eq. (\ref{mean num}) is that the first term is the spontaneous production
while the second term is the stimulated emission in the presence of in-vacuum particles.
In the specific case of the Bunch-Davies vacuum,
the probability for particle production is the Gibbons-Hawking radiation of bosons\cite{gibbons-hawking}
\begin{eqnarray}
P_k =  \frac{1}{e^{2 \pi p} -1}.
\label{gibbons-hawking}
\end{eqnarray}
The particle production is independent of the momentum and the scale factor $a(t)$
and is determined by the mass and the Hubble constant.

\section{Gaussian Wave Packets} \label{gaussian}

We shall use the Schr\"{o}dinger picture since quantum states are
c-number wave functions that carry the same information as the Heisenberg field operator.
Though not considered in this paper, the Schr\"{o}dinger picture is also convenient for a nonlinear theory
of self-interactions, modulo the renormalization of the wave function, the vacuum energy and the coupling constants.

The time-dependent Hamiltonian (\ref{ham}) in an expanding universe may give an adverse feeling for the Schr\"{o}dinger
picture
\begin{eqnarray}
i \frac{\partial}{\partial t} \Psi_k (\phi_k, t) = \hat{H}_k (t) \Psi_k (\phi_k, t).\label{sch eq}
\end{eqnarray}
However, it has been known for a long time that the Gaussian wave function could be expressed in terms of a classical solution
to eq. (\ref{mod eq}), for instance, see ref. [\refcite{guth-pi}]. It may be understood from the fact that
the Heisenberg field operator also satisfies the same equation (\ref{mod eq}) and the Heisenberg picture provides
the same quantum information as the Schr\"{o}dinger picture.
In fact, the Gaussian wave function on a spacelike hypersurface $\Sigma_t$ of constant $t$
for each Fourier mode, which is the solution to the Schr\"{o}dinger (\ref{sch eq}) and
is normalized to unity, is given by\cite{kim-page}
\begin{eqnarray}
\Psi_{k} (\phi_k, t) = \Bigl( \frac{\varphi_k}{\sqrt{2 \pi} |\varphi_k|^2} \Bigr)^{1/2}
\exp \Bigl( \frac{ia^3}{2} \frac{\dot{\varphi}^*_k}{\varphi^*_k} \phi_k^2 \Bigr), \label{gauss wave}
\end{eqnarray}
where $\varphi_k$ is a complex solution to eq. (\ref{mod eq}) that satisfies the quantization rule (\ref{quant rule}).

In the invariant-operator picture by Lewis and Riesenfeld, the Hilbert space for a time-dependent Hamiltonian of
quadratic order can be constructed exactly in the same manner as
for a time-independent Hamiltonian. In this picture we look
for the operators that satisfy the Liouville-von Neumann equation
\begin{eqnarray}
i \frac{\partial}{\partial t} \hat{b}_k (t) + [\hat{b}_k (t) \hat{H}_k (t)] = 0. \label{in op}
\end{eqnarray}
Note that eq. (\ref{in op}) is the equation for the density operator and describes the backward evolution
\begin{eqnarray}
\hat{b}_k^{\rm (I)} (t) = \hat{U}_k^{\dagger} (t) \hat{b}_k^{\rm (S)} \hat{U}_k (t), \label{lv op}
\end{eqnarray}
in contrary to the forward evolution of the Heisenberg operator
\begin{eqnarray}
\hat{b}_k^{\rm (H)} (t) = \hat{U}_k (t) \hat{b}_k^{\rm (S)} \hat{U}_k^{\dagger} (t),
\end{eqnarray}
where $\hat{U}_k$ is the evolution operator
\begin{eqnarray}
i \frac{\partial}{\partial t} \hat{U}_k (t) = \hat{H}_k (t) \hat{U}_k (t).\label{ev op}
\end{eqnarray}
Indeed the operator (\ref{lv op}) satisfies eq. (\ref{in op}) with respect to the Hamiltonian
\begin{eqnarray}
\hat{H}_k^{\rm (I)} (t) = \hat{U}_k^{\dagger} (t) \hat{H}_k^{\rm (S)} \hat{U}_k (t).
\end{eqnarray}

For the quadratic Hamiltonian we can directly find the invariant operators
(the superscript (I) dropped below) for eq. (\ref{in op}), which are
first order in the momentum and the position operators,\cite{mmt,bfv,kim-page,kim-kim,pfr,gds}
\begin{eqnarray}
\hat{b}_k (t) &=& i \bigl( \varphi_k^* (t) \hat{\pi}_k - a^3 (t) \dot{\varphi}_k^* (t) \hat{\phi}_k \bigr), \nonumber\\
\hat{b}_k^{\dagger} (t) &=& - i \bigl( \varphi_k (t) \hat{\pi}_k - a^3 (t) \dot{\varphi}_k (t) \hat{\phi}_k \bigr).
\end{eqnarray}
In the above $\hat{\pi}_k$ and $\hat{\phi}_k$ are the Schr\"{o}dinger operators and $\varphi_k (t)$ is a complex solution
to eq. (\ref{mod eq}) together with the quantization rule (\ref{quant rule}). Remarkably, these operators
satisfy the equal-time commutation relation
\begin{eqnarray}
[\hat{b}_k (t), \hat{b}_k^{\dagger} (t) ] = 1,
\end{eqnarray}
and, moreover, play the role of the time-dependent annihilation and creation operators since
they reduce to the standard ones for a time-independent Hamiltonian, modulo time-dependent phase factors.
In fact, the Gaussian wave function (\ref{gauss wave}) is annihilated by $\hat{b}_k (t)$.

\section{Quantum Dynamics} \label{quantum dyn}

The Hamiltonian (\ref{ham}) has the algebra SU(1,1), for which we may choose a time-independent Hermitian basis
\begin{eqnarray}
\hat{L}^{(-)} = \frac{1}{2} \hat{\phi}_k^2 , \quad \hat{L}^{(+)} = \frac{1}{2} \hat{\pi}_k^2, \quad \hat{L}^{(0)} = \frac{1}{2} \bigl( \hat{\phi}_k \hat{\pi}_k + \hat{\pi}_k \hat{\phi}_k \bigr). \label{quad}
\end{eqnarray}
As the field and the conjugate momentum operators have the expression
\begin{eqnarray}
\hat{\phi}_k (t) = \varphi_k (t) \hat{b}_k (t) + \varphi_k^* (t) \hat{b}_k (t), \quad
\hat{\pi}_k (t) = a^3 (t) \bigl[ \dot{\varphi}_k (t) \hat{b}_k (t) + \dot{\varphi}_k^* (t) \hat{b}_k (t) \bigr], \label{field op}
\end{eqnarray}
we may use another Hermitian basis\cite{kim-schubert}
\begin{eqnarray}
\hat{M}_k^{(0)} (t) &=& \hat{b}_k^{\dagger} (t) \hat{b}_k (t) + \hat{b}_k (t) \hat{b}_k^{\dagger} (t), \label{0 op}\\
\hat{M}_k^{(+)} (t) &=& \hat{b}_k^2 (t) + \hat{b}_k^{\dagger 2} (t), \quad
\hat{M}_k^{(-)} (t) = i \bigl[ \hat{b}_k^2 (t) - \hat{b}_k^{\dagger 2} (t) \bigr]. \label{- op}
\end{eqnarray}
This basis has the SU(1,1) group structure
\begin{eqnarray}
[ \hat{M}_k^{(0)} (t), \hat{M}_{k'}^{(\pm)} (t) ] =  \pm 2 i \hat{M}_k^{(\mp)} (t) \delta_{kk'}, \,\,
[ \hat{M}_k^{(+)} (t), \hat{M}_{k'}^{(-)} (t) ] = - 2 i \hat{M}_k^{(0)} (t) \delta_{kk'}.
\end{eqnarray}
Note that $\hat{M}_k^{(0)} (t)$ is the number operator on the hypersurface $\Sigma_t$.

Using eq. (\ref{field op}) we express the quadratic operators for the field and the conjugate momentum in terms of
the basis (\ref{0 op}, \ref{- op})
\begin{eqnarray}
\hat{\phi}_k^2 &=& |\varphi_k (t)|^2  \hat{M}_k^{(0)} (t) + \frac{1}{2} \bigl( \varphi_k^{*2} (t) + \varphi_k^{2} (t) \bigr) \hat{M}_k^{(+)} (t)
\nonumber\\ && + \frac{i}{2}\bigl( \varphi_k^{*2} (t) - \varphi_k^{2} (t) \bigr) \hat{M}_k^{(-)} (t), \nonumber\\
\hat{\pi}_k^2 &=& a^6 (t) \Bigl[ |\dot{\varphi}_k (t)|^2  \hat{M}_k^{(0)} (t) + \frac{1}{2} \bigl( \dot{\varphi}_k^{*2} (t) +
\dot{\varphi}_k^{2} (t) \bigr) \hat{M}_k^{(+)} (t)
\nonumber\\ && + \frac{i}{2} \bigl( \dot{\varphi}_k^{*2} (t) - \dot{\varphi}_k^{2} (t) \bigr) \hat{M}_k^{(-)} (t) \Bigr], \nonumber\\
\hat{\phi}_k \hat{\pi}_k + \hat{
\pi}_k \hat{\phi}_k &=& a^3 (t) \Bigl[ (|\varphi_k (t)|^2)^{\cdot}  \hat{M}_k^{(0)} (t) + \frac{1}{2} \bigl( (\varphi_k^{*2} (t))^{\cdot} +
(\varphi_k^2 (t))^{\cdot} \bigr) \hat{M}_k^{(+)} (t)
\nonumber\\ && + \frac{i}{2} \bigl( (\varphi_k^{*2} (t))^{\cdot} -
(\varphi_k^2 (t))^{\cdot} \bigr) \hat{M}_k^{(-)} (t) \Bigr].
\end{eqnarray}
Conversely, the number operator can be expressed in the basis (\ref{quad}) as
\begin{eqnarray}
\frac{1}{2} \hat{M}_k^{(0)} (t) = |\varphi_k (t)|^2  \hat{\pi}_k^2 + a^6(t) |\dot{\varphi}_k (t)|^2  \hat{\phi}_k^2
- a^3 (t) (|\varphi_k (t)|^2)^{\cdot} \frac{1}{2} \bigl( \hat{\phi}_k \hat{\pi}_k + \hat{\pi}_k \hat{\phi}_k \bigr).
\label{num rel}
\end{eqnarray}
The quadratic variances with respect to the Gaussian wave function (\ref{gauss wave}) on $\Sigma_t$ take the values
\begin{eqnarray}
\langle \Psi_k (t) \vert \hat{\phi}_k^2 \vert \Psi_k (t) \rangle = |\varphi_k (t)|^2, \quad
&& \langle \Psi_k (t) \vert \hat{\pi}_k^2 \vert \Psi_k (t) \rangle = a^6 (t) |\dot{\varphi}_k (t)|^2, \nonumber\\
\langle \Psi_k (t) \vert  \hat{\phi}_k \hat{\pi}_k + \hat{\pi}_k \hat{\phi}_k \vert \Psi_k (t) \rangle &=&
a^3 (t) (|\varphi_k (t)|^2)^{\cdot}. \label{quad exp}
\end{eqnarray}

\section{Particle Production in Quantum Dynamics} \label{particle prod}

We may raise the question whether the operator (\ref{0 op}) indeed provides a proper definition for the particle number
\begin{eqnarray}
\hat{N}_k^{\dagger} (t) =  \hat{b}_k^{\dagger} (t) \hat{b}_k (t) = \frac{1}{2} \hat{M}_k^{(0)} (t) - \frac{1}{2}. \label{num op}
\end{eqnarray}
The Gaussian wave function (\ref{gauss wave}) has the zero particle number on $\Sigma_t$
\begin{eqnarray}
\langle \Psi_k (t) \vert \hat{N}_k (t) \vert \Psi_k (t) \rangle = 0.
\end{eqnarray}
Thus the number operator $\hat{N}_k^{\dagger} (t)$ measures the amount of particles carried by the Gaussian wave function
on each $\Sigma_t$ through the evolution, which is a conserved quantum number, that is, the harmonic excitation number.
We use the Gaussian wave function on the hypersurface $\Sigma_{- \infty}$ in the past infinity
as the in-vacuum state for each $k$ and define the in-vacuum as the tensor product
\begin{eqnarray}
\vert {\rm vac, in} \rangle = \prod_k \vert \Psi_k (- \infty) \rangle.
\end{eqnarray}
Similarly we define the out-vacuum as
\begin{eqnarray}
\vert {\rm vac, out} \rangle = \prod_k \vert \Psi_k ( \infty) \rangle.
\end{eqnarray}
Then the number of in-vacuum particles contained in the Gaussian wave function on $\Sigma_t$ or the number of out-vacuum particles
contained in the in-vacuum Gaussian wave function on $\Sigma_{- \infty}$ is
\begin{eqnarray}
n_k (t) = \langle \Psi_k (t) \vert \hat{N}_k (- \infty) \vert \Psi_k (t) \rangle = \langle \Psi_k (- \infty) \vert \hat{N}_k (t) \vert \Psi_k (- \infty) \rangle.
\end{eqnarray}
The second equality is a consequence of the reciprocality, as will be explicitly shown below.

Taking the expectation value of the operator $\hat{M}_k^{(0)} (- \infty)$ in eq. (\ref{num rel})
with respect to the Gaussian wave function on $\Sigma_{t}$, we obtain
\begin{eqnarray}
n_k (t) &=& a^6 (t) |\varphi_k (- \infty)|^2  |\dot{\varphi}_k (t)|^2 + a^6(- \infty) |\dot{\varphi}_k (- \infty)|^2  |\varphi_k (t)|^2
\nonumber\\&& - \frac{1}{2} a^3 (- \infty) a^3 (t) (|\varphi_k (- \infty)|^2)^{\cdot} (|\varphi_k (t)|^2)^{\cdot} -\frac{1}{2}.
\label{par num}
\end{eqnarray}
We find the Bogoliubov transformation between $\Sigma_{t_0}$ and $\Sigma_t$
\begin{eqnarray}
\hat{a}_k (t_0) = \mu_k (t_0, t) \hat{a}_k (t) + \nu_k (t_0, t) \hat{a}^{\dagger}_k (t), \nonumber\\
\hat{a}^{\dagger}_k (t_0) = \mu_k^* (t_0, t) \hat{a}^{\dagger}_k (t) + \nu_k^* (t_0, t) \hat{a}_k (t), \label{bog tran1}
\end{eqnarray}
where
\begin{eqnarray}
\mu_k (t_0, t) &=& i [a^3 (t) \varphi_k^* (t_0) \dot{\varphi}_k (t) - a^3 (t_0) \varphi_k (t) \dot{\varphi}_k^* (t_0)], \nonumber\\
\nu_k (t_0, t) &=& i [a^3 (t) \varphi_k^* (t_0) \dot{\varphi}_k^* (t) - a^3 (t_0) \varphi_k^* (t) \dot{\varphi}_k^* (t_0)]. \label{bog coef2}
\end{eqnarray}
The Bogoliubov transformation (\ref{bog tran1}) is symmetric under the interchange of
$\hat{a}_k (t_0)$, $\hat{a}^{\dagger}_k (t_0)$ and $\hat{a}_k (t)$, $\hat{a}^{\dagger}_k (t)$, so
the inverse transformation
\begin{eqnarray}
\hat{a}_k (t) = \mu_k (t, t_0) \hat{a}_k (t_0) + \nu_k (t, t_0) \hat{a}^{\dagger}_k (t_0), \nonumber\\
\hat{a}^{\dagger}_k (t) = \mu_k^* (t, t_0) \hat{a}^{\dagger}_k (t_0) + \nu_k^* (t, t_0) \hat{a}_k (t_0),
\end{eqnarray}
has the coefficients
\begin{eqnarray}
\mu_k (t, t_0) = \mu_k^* (t_0, t), \quad \nu_k (t, t_0) = - \nu_k (t_0, t). \label{inv bog coef}
\end{eqnarray}
It can be shown that the particle number is given by
\begin{eqnarray}
n_k (t) = |\nu_k (- \infty, t)|^2 = |\nu_k (t, - \infty)|^2.
\end{eqnarray}
Furthermore, the Bogoliubov relation holds
\begin{eqnarray}
|\mu_k (t_0, t)|^2 - |\nu_k (t_0, t)|^2 = 1.
\end{eqnarray}
However, the Bogoliubov coefficients (\ref{bog coef1}) cannot be obtained from those
(\ref{bog coef2}) using the asymptotic solutions in section \ref{dS radiation}.

\section{Vacuum Persistence Amplitude} \label{vac per}

One important concept for the particle production and the vacuum instability is
the vacuum persistence amplitude introduced by Schwinger
and DeWitt,\cite{schwinger,dewitt}
which is the scattering-matrix ($S$-matrix) amplitude between two different hypersurfaces
\begin{eqnarray}
e^{i W_k (t, t_0)} = \langle \Psi_k (t) \vert \Psi_k (t_0) \rangle. \label{s-mat}
\end{eqnarray}
In particle physics the $S$-matrix takes two asymptotic limits $t_0 = - \infty$ and $t= \infty$, that is,
before and after interactions.
But it is still legitimate to use the definition (\ref{s-mat}) in an intermediate region, though the physical meaning
requires further clarification and investigation.
The integration of two Gaussian wave functions (\ref{gauss wave}) gives
\begin{eqnarray}
W_k (t, t_0) =  i \ln \bigl( \mu_k (t_0, t) \bigr) = i \ln \bigl( \mu_k^* (t, t_0) \bigr),
\end{eqnarray}
where $\mu_k$ is given in eqs. (\ref{bog coef2}, \ref{inv bog coef}) and the mode $k$ counts both ${\bf k}$ and $- {\bf k}$.

The vacuum persistence amplitude for the field is the product of eq. (\ref{s-mat})
\begin{eqnarray}
e^{i W (t, t_0)} = \prod_{k} \langle \Psi_k (t) \vert \Psi_k (t_0) \rangle,
\end{eqnarray}
and thereby the effective action is
\begin{eqnarray}
W (t, t_0) =  i {\cal V} (t) \int d^3 {\bf k} \ln \bigl( \mu_{\bf k} (t_0, t) \bigr) =
i {\cal V} (t) \int d^3 {\bf k} \ln \bigl( \mu^*_{\bf k} (t, t_0) \bigr). \label{eff ac}
\end{eqnarray}
Here $\mu_{\bf k} (t_0, t)$ is the Bogoliubov coefficient obtained by substituting ${\bf k}$ for $k$ in eq. (\ref{bog coef2})
and ${\cal V} (t)$ is a spatial volume relevant to the
hypersurface on $\Sigma_t$ and may be chosen as ${\cal V} (t) = a^3 (t) \int d^3 {\bf x}$.
The effective action (\ref{eff ac}) involves the divergent terms that renormalize the vacuum energy or
the mass, and the gravitational constant or the charge.
In the in-out formalism equation eq. (\ref{eff ac}) has been used to find the effective action for the massive scalar field
in the global geometry of dS space\cite{kim2010}
and in quantum electrodynamics (QED) the effective action in a time-dependent electric field.\cite{kly2008}

The vacuum persistence (twice of the imaginary part) relates to the particle number through the relation
\begin{eqnarray}
2\, {\rm Im} (W_k (t, t_0)) = \ln \bigl( |\mu_k (t, t_0)|^2 \bigr) = \ln \bigl( 1 + |\nu_k (t, t_0)|^2 \bigr).
\end{eqnarray}
And the probability for the in-state on $\Sigma_{t_0}$ to remain in the out-state on $\Sigma_t$ is
\begin{eqnarray}
|\langle 0, t \vert 0, t_0 \rangle|^2 = e^{-2{\rm Im} (W (t, t_0))} =
\exp \Bigl[- {\cal V} (t) \int d^3 {\bf k} \ln \bigl( 1 + n_{\bf k} (t, t_0)|^2 \bigr) \Bigr]. \label{in-out prob}
\end{eqnarray}
Thus the vacuum of dS space becomes unstable due to the dS radiation.
It is shown that the number of in-vacuum particles carried by the Gaussian wave function on at a later time hypersurface
$\Sigma_t$ is dominated by the first term in eq. (\ref{par num})
\begin{eqnarray}
n_k (t) \simeq a^3 (t) \bigl( a^3 (t) |\dot{\varphi}_k (t)|^2 \bigr) |\dot{\varphi}_k (t)|^2 |\varphi_k (- \infty)|^2
\end{eqnarray}
and is enhanced by the exponential scale factor and by the infinitely blueshifted wavelength on the initial surface
while the parenthesis is finite.
This catastrophic production of in-vacuum particles in the infinity future is a generic phenomenon in an expanding
universe with a cosmological singularity.
An interesting observation is that if a regularization scheme removes $a^3 (t)$ and $|\varphi_k (- \infty)|^2$
in the vacuum persistence probability (\ref{in-out prob}), the remaining term is
related to the Gibbons-Hawking radiation.

\section{Wave Packet for Out-Vacuum} \label{wave out}

The out-vacuum solution,
\begin{eqnarray}
\varphi^{{\rm (out)}}_k (t) = \Bigl( \frac{\pi }{2H \sinh (\pi p)} \Bigr)^{1/2} e^{-3 H t/2} J_{ip} (z), \label{out sol2}
\end{eqnarray}
satisfies the quantization rule (\ref{quant rule}) and has the same time-dependent factor
as eq. (\ref{asym sol2}) up to a constant phase factor. The corresponding Gaussian wave function
\begin{eqnarray}
\Psi^{{\rm (out)}}_k (\phi_k, t) =
\Bigl( \frac{\varphi^{{\rm (out)}}_k}{\sqrt{2\pi} |\varphi^{{\rm (out)}}_k|^2} \Bigr)^{1/2}
\exp \Bigl(\frac{ia^3}{2} \frac{\dot{\varphi}^{{\rm (out)}*}_k}{\varphi^{{\rm (out)}*}_k} \phi_k^2 \Bigr), \label{out wave}
\end{eqnarray}
is another exact solution to the Schr\"{o}dinger equation (\ref{sch eq}) since eq. (\ref{out sol2})
is a complex solution to eq. (\ref{mod eq}). The Bunch-Davies vacuum solution (\ref{gen sol})
($c^{(1)}_k = 1$, $c^{(2)}_k = 0$) now has the Bogoliubov transformation
\begin{eqnarray}
\varphi_k (t) = \alpha_k \varphi^{{\rm (out)}}_k (t) + \beta_k \varphi^{{\rm (out)}*}_k (t),
\end{eqnarray}
where
\begin{eqnarray}
\alpha_k = \Bigl(\frac{e^{\pi p}}{2\sinh(\pi p)} \Bigr)^{1/2}, \quad \beta_k = - \Bigl(\frac{e^{- \pi p}}{2\sinh(\pi p)} \Bigr)^{1/2}.
\end{eqnarray}
The particle production $|\beta_k|^2$ is the Gibbons-Hawking radiation for massive bosons ({\ref{gibbons-hawking}).

In analogy with the vacuum persistence amplitude between the in-vacuum and the out-vacuum in section \ref{vac per},
we may compute the matrix amplitude on the same hypersurface $\Sigma_t$
\begin{eqnarray}
e^{i W^{\rm (out)}_k (t)} = \langle \Psi_k (t) \vert \Psi^{\rm (out)}_k (t) \rangle. \label{out-mat}
\end{eqnarray}
Then the vacuum persistence (twice of the imaginary part) relates to the particle number through the relation
\begin{eqnarray}
2\, {\rm Im} (W^{\rm (out)}_k (t)) = \ln \bigl( |\alpha_k|^2 \bigr) = \ln \bigl( 1 + |\beta_k|^2 \bigr),
\end{eqnarray}
and explains the Gibbons-Hawking radiation.

\section{Conclusion}

In this paper we have studied the production of massive particles in a dS space using the quantum dynamics
that unifies the invariant-operator picture with the in-out formalism.
In the quantum dynamics a massive field in an expanding FRW universe or a dS space
is equivalent to an infinite sum of time-dependent oscillators
in which the time-dependent mass comes from the expanding spatial volume and the frequencies
from the redshifted or blueshifted wavelengths. The main advantage of the invariant-operator picture is that there exist
the time-dependent annihilation and creation
operators which generate the exact quantum states, the Gaussian wave function being the simplest one among them.
The Gaussian wave function provides the in-vacuum in the past infinity and the out-vacuum in the future infinity.
Hence the time-dependent annihilation and creation operators connect these two asymptotic regions
through the Bogoliubov transformation.

We have found the Gaussian wave packets corresponding to the Bunch-Davies vacuum and its one-parameter family.
The scattering-matrix of Gaussian wave packets between two different spacelike
hypersurfaces directly gives the vacuum persistence amplitude,
which is equivalent to that from the Bogoliubov transformation and its coefficients.
The vacuum persistence amplitude is an important tool in quantifying how the massive field
probes the background dS space in analogy with QED in a strong electric field background.
However, in the planar coordinates of dS space the number of in-vacuum particles
carried by the Gaussian wave packet
exponentially increases in proportion to the expanding spatial volume, in contrast with
the Gibbons-Hawking radiation. This is because the momentum variance, the leading term
for particle number, exponentially increases in the infinity future
while the wave packet is very sharply peaked in the field.
On the other hand, the matrix amplitude between the Gaussian wave packet and the out-vacuum Gaussian
wave packet on the same hypersurface in the far future gives the vacuum persistence that explains
the Gibbons-Hawking radiation.
This raises a fundamental question which vacuum persistence amplitude is the proper definition for
the effective action and the dS radiation. The related issues including
the massless scalar field will be addressed in a future publication.

\section*{Acknowledgments}
The author thanks Bo-Qiang Ma for the warmest hospitality during CosPA 2011, Beijing University, China,
October 28-31, 2011. He also thanks W-Y.~Pauchy Hwang for the warmest hospitality at
National Taiwan University, where this paper was completed, and for his continuing efforts for
the Asia Pacific Organization of Cosmology and Particle Astrophysics (APCosPA).
The participation of the CosPA symposium was supported in part
by Basic Science Research Program through the National Research Foundation of Korea (NRF) funded by the Ministry of Education, Science and Technology (2011-0002-520) and in part by Beijing University.
The work was supported in part by National Science Council Grant (NSC 100-2811-M-002-188).

%\begin{thebibliography}{000} %for 3 digits
%\begin{thebibliography}{00}  %for 2 digits

\end{document}